\documentclass[aps,prd,notitlepage,nofootinbib]{revtex4-1}
\pdfoutput=1
\usepackage[top=3cm, bottom=3cm, left=2cm, right=2cm]{geometry}
\usepackage[usenames,dvipsnames,svgnames]{xcolor}
\definecolor{darkblue}{rgb}{0.0,0.1,0.3} 
\definecolor{darkgreen}{rgb}{0,0.65,0}
\definecolor{dblue4}{rgb}{0.06,0.31,0.55} 
\definecolor{nicered}{rgb}{0.7,0.1,0.1}
\definecolor{nicegreen}{rgb}{0.1,0.5,0.1}

\usepackage[utf8]{inputenc}
\usepackage{amsmath,amssymb,amsfonts,amsthm}
\usepackage{tabularx}
\usepackage{multirow}
\usepackage{dsfont}
\newcolumntype{Y}{>{\centering\arraybackslash}X}

\usepackage{xcolor}
\usepackage[colorlinks=true,citecolor= nicegreen,linkcolor=nicered]{hyperref}

\usepackage{graphicx}
\graphicspath{{figures/}}
\usepackage{cancel}

\usepackage[colorinlistoftodos]{todonotes}
\usepackage{soul}
\usepackage{comment}
\includecomment{details}
\specialcomment{details}
{\begingroup}{\endgroup}
\excludecomment{details}
\usepackage{tikz}

\newcommand{\UPTC}{Escuela de Física, Universidad Pedagógica y Tecnológica de Colombia,\\
Avenida Central del Norte \# 39-115, Tunja, Colombia}
\newcommand{\UdeA}{Instituto de Física, Universidad de Antioquia,\\Calle 70 \# 52-21, Apartado Aéreo 1226, Medellín, Colombia}

\begin{document}
\title{Minimal radiative Dirac neutrino mass models}

\author{Julian Calle~${}^1$} \email{julian.callem@udea.edu.co} 
\author{Diego Restrepo~${}^1$}\email{restrepo@udea.edu.co}
\author{Carlos E. Yaguna~${}^2$}\email{carlos.yaguna@uptc.edu.co}
\author{Óscar Zapata~${}^1$}\email{oalberto.zapata@udea.edu.co}

\affiliation{$^1$~\UdeA}
\affiliation{$^2$~\UPTC}

\begin{abstract}
  Neutrinos may be Dirac particles whose masses arise radiatively at one-loop, naturally explaining their small values. In this work we show that all the one-loop realizations of the dimension-five operator to effectively generate Dirac neutrino masses can be implemented by using a single local symmetry: $\operatorname{U}(1)_{B-L}$. Since this symmetry is anomalous, new chiral fermions, charged under $B-L$, are required. The minimal model consistent with neutrino data includes three chiral fermions, two of them with the same lepton number. The next minimal models contain five chiral fermions and their  $B-L$ charges can be fixed by requiring  a dark matter candidate in the spectrum.  We list the full particle content as well as the relevant Lagrangian terms for each of these models. They are new and simple models that can simultaneously accommodate  Dirac neutrino masses (at one-loop) and dark matter without invoking any discrete symmetries. 
\end{abstract}

\maketitle

\section{Introduction}
\label{sec:intro}

The interpretation of neutrino experimental data in terms of neutrino
oscillations is compatible with both Majorana or Dirac neutrino
masses \cite{Tanabashi:2018oca}. The former possibility  has received the most attention but, given the lack of signals in neutrinoless double beta decay experiments
\cite{KamLAND-Zen:2016pfg,Agostini:2018tnm,Aalseth:2017btx,Alduino:2017ehq,Albert:2017owj,Arnold:2016bed}, the latter cannot be dismissed. If neutrinos are Dirac particles, the Standard Model (SM) particle content must be extended with right-handed neutrinos, and some symmetry must be imposed to prevent their Majorana mass terms. At least a $Z_3$ symmetry is required to guarantee the neutrino Diracness through
\begin{align}
  \label{eq:tld}
  \mathcal{L}_{\nu}=&\,y_D\left( \nu_R \right)^{\dagger} L\cdot H +\text{h.c.} \,,
\end{align}
with\footnote{Throughout the text we will follow
  the convention of defining only left-handed Weyl spinors~\cite{Dreiner:2008tw}, and we will use the
  $\operatorname{SU}(2)$ metric to build scalar products.} $L\cdot H=\epsilon_{ab}L^a H^b$, where $L$ is the lepton doublet, $H$ is the
SM Higgs doublet with hypercharge $Y=1$, and $y_D$ is the matrix of neutrino Yukawa couplings.  To be compatible with  neutrino
oscillation data \cite{deSalas:2017kay}, $y_D$ should be at least of order $2\times 3$.  
A possible assignment for the set of SM fields that transform
non-trivially under $Z_3$ is: $L\sim \omega$,
$\left( e_R \right)^{\dagger} \sim \omega^2$ and
$\left( \nu_R \right)^{\dagger} \sim \omega^2$, with $\omega^3=1$.
At this level, the neutrino mass problem is not longer a
phenomenological issue but a theoretical one, in which it is necessary
to explain the smallness of the Yukawa couplings in $y_D$, which must be  of order
$10^{-11}$.

To do so, we assume that the symmetry allows for the 5-dimensional operator with total lepton number conservation~\cite{Gu:2006dc}
\begin{align}
  \label{eq:blop}
  \mathcal{L}_5=
  \frac{h}{\Lambda} \left( \nu_R \right)^\dagger  L\cdot H\, S^{*} + 
\text{h.c.}\,,
\end{align}
where $\Lambda$ is the new physics scale, and that this operator
is first realized at one-loop level~\cite{Yao:2018ekp}
(see~\cite{Roncadelli:1983ty,Roy:1983be,Gu:2007mc,Ma:2014qra,Ma:2016mwh,Yao:2018ekp,CentellesChulia:2018gwr,CentellesChulia:2018bkz,Bernal:2018aon} for the tree-level realizations).  

Regarding the symmetry, we follow the usual approach of promoting baryon number ($B$) minus lepton number  ($L$) from an  accidental global symmetry of the SM, to  a local Abelian symmetry, $\operatorname{U}(1)_{B-L}$, which is spontaneously broken. One of the main novelties of our work is that we do not impose any other symmetries, discrete or otherwise. Thus, the charges of the right-handed neutrinos under $\operatorname{U}(1)_{B-L}$ should be such that the tree-level Dirac mass term~\eqref{eq:tld} is forbidden. This requirement automatically excludes the usual assignment where the three right-handed neutrinos have $B-L$ charges equal to  $-1$.

The classification of all the topologies at one-loop level that realize the effective operator \eqref{eq:blop} has been presented in~\cite{Yao:2018ekp}.
There, in addition to the $\operatorname{U}(1)_{B-L}$ symmetry, at
least one additional $Z_2$ symmetry was imposed to avoid the Majorana
mass terms for the right-handed neutrinos, and a further $Z_2'$ was
required to avoid $i)$ the appearance of tree-level realizations
in the cases when $\operatorname{U}(1)_{B-L}$ is not able to do it,
and $ii)$ to have a dark matter candidate in the particle spectrum --one of the new
particles needed to realize (\ref{eq:blop}).
Here, we focus instead on the simplest realizations of each topology that can
be realized with a single symmetry, $\operatorname{U}(1)_{B-L}$. This same symmetry would be responsible for the stability of possible dark matter candidates 
appearing in the different realizations. Let us stress that until now in the literature there is not a simple realization of operator~\eqref{eq:blop} at one-loop invoking
only a single symmetry. There were some efforts in this direction but either some Majorana
terms were left out, which would need to be forbidden with an extra
$Z_2$-symmetry~\cite{Wang:2017mcy,Han:2018zcn}, or the found models
require many extra fields~\cite{Wang:2017mcy}.

Using only $\operatorname{U}(1)_{B-L}$, we find, for each topology,  the realization with the minimum number of fields. The minimal model requires three chiral fields, two of them  sharing the same lepton number, so that the spectrum contains two massive Dirac neutrinos. The next minimal models include five chiral fields. Interestingly, their $B-L$ charges are fixed once the requirement to have a dark matter particle is imposed. Hence, these new models may account for dark matter and Dirac neutrino masses (at one-loop) without invoking any discrete symmetries.

The rest of the paper is organized as follows. In the next section we introduce the notation and derive the conditions necessary to realize the different topologies that give rise to one-loop Dirac neutrino masses. Our main results are presented in section \ref{sec:sol}. There, the particle content of the minimal models, with three and five chiral fields, is spelled out. Finally, in section \ref{sec:con} our conclusions are drawn.

\section{General setup}
\label{sec:setup}
\begin{figure}
  \centering
  \includegraphics[scale=0.55]{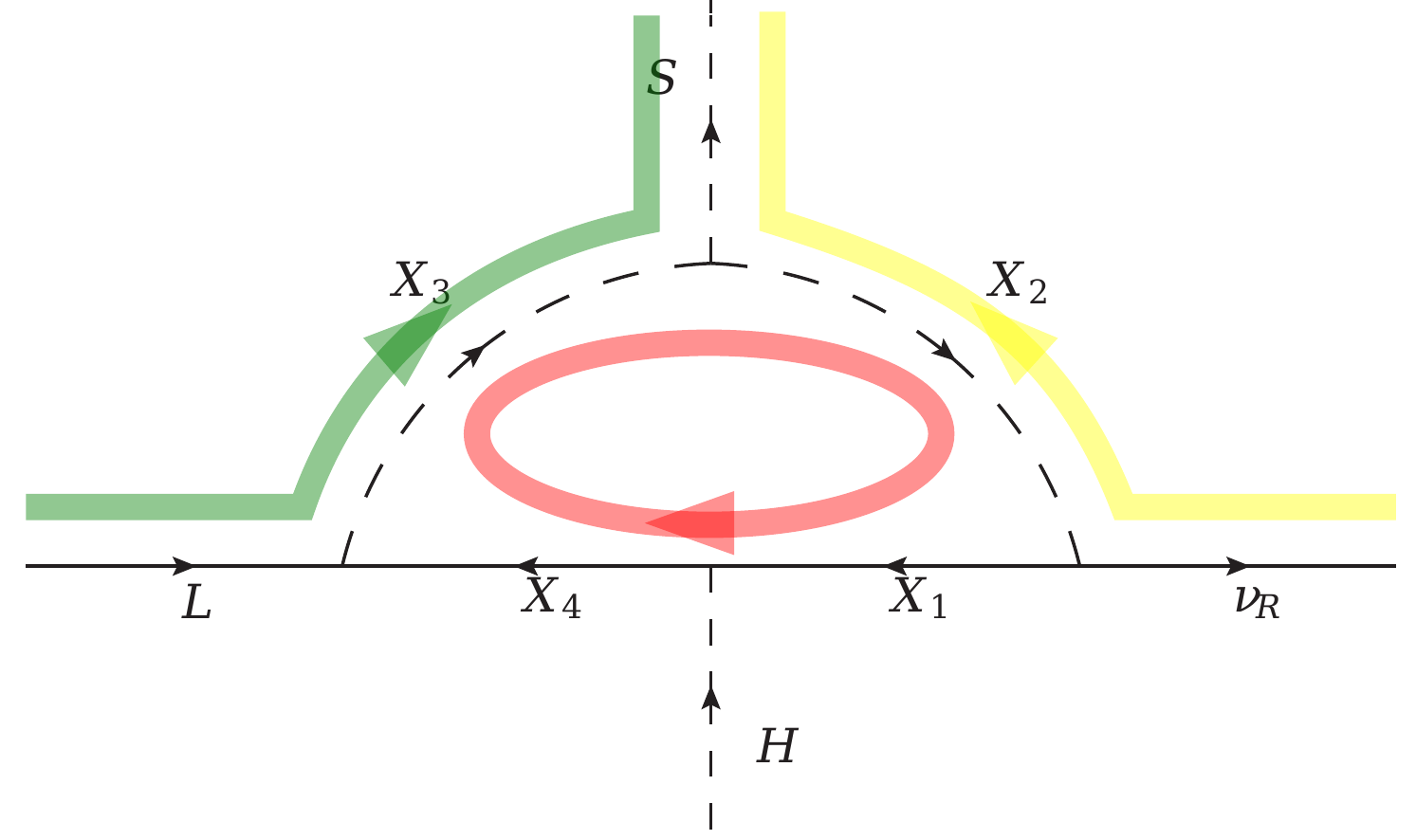}
  \includegraphics[scale=0.55]{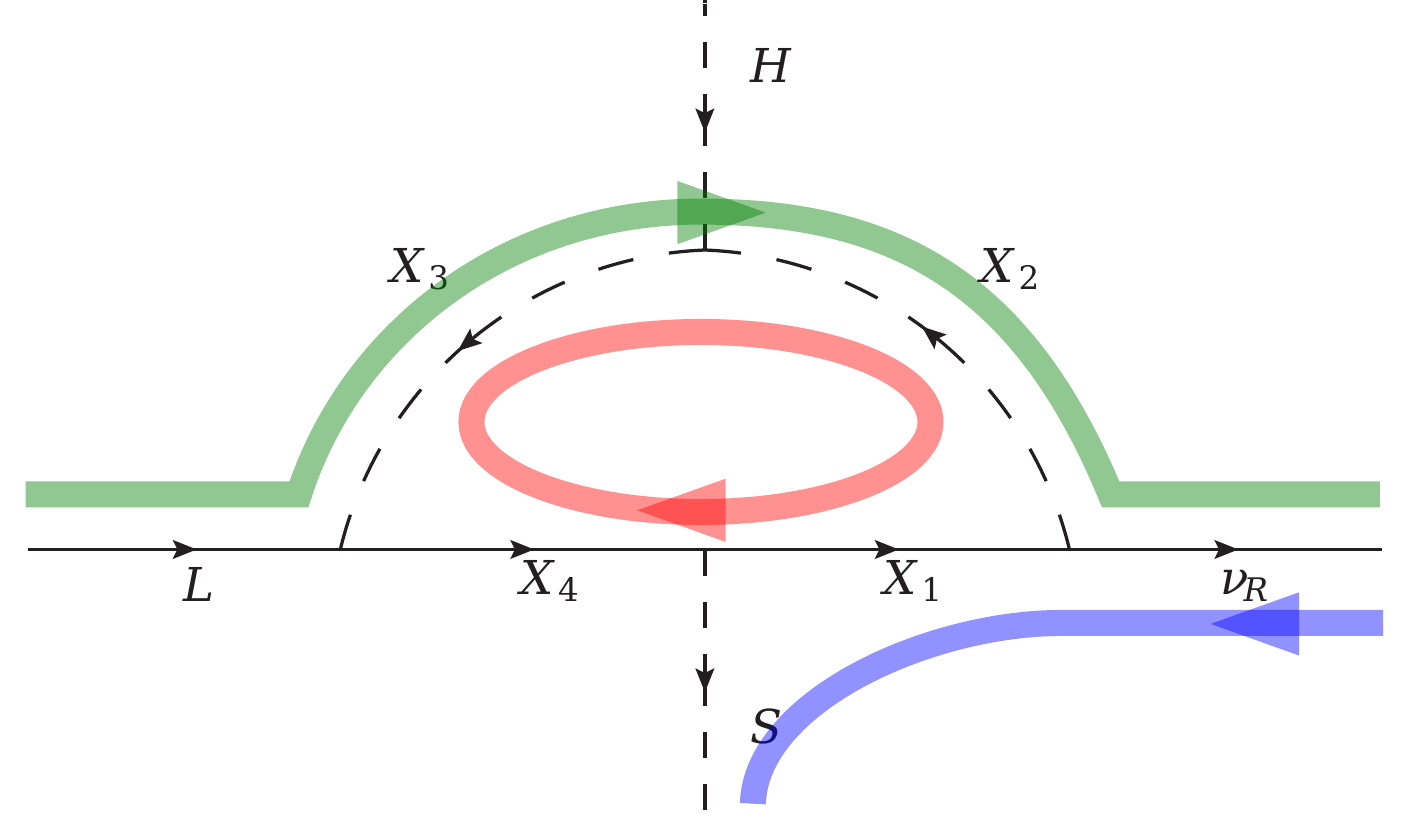}\\
  T1-3-D \hspace{4cm} T1-3-E\\
    \includegraphics[scale=0.55]{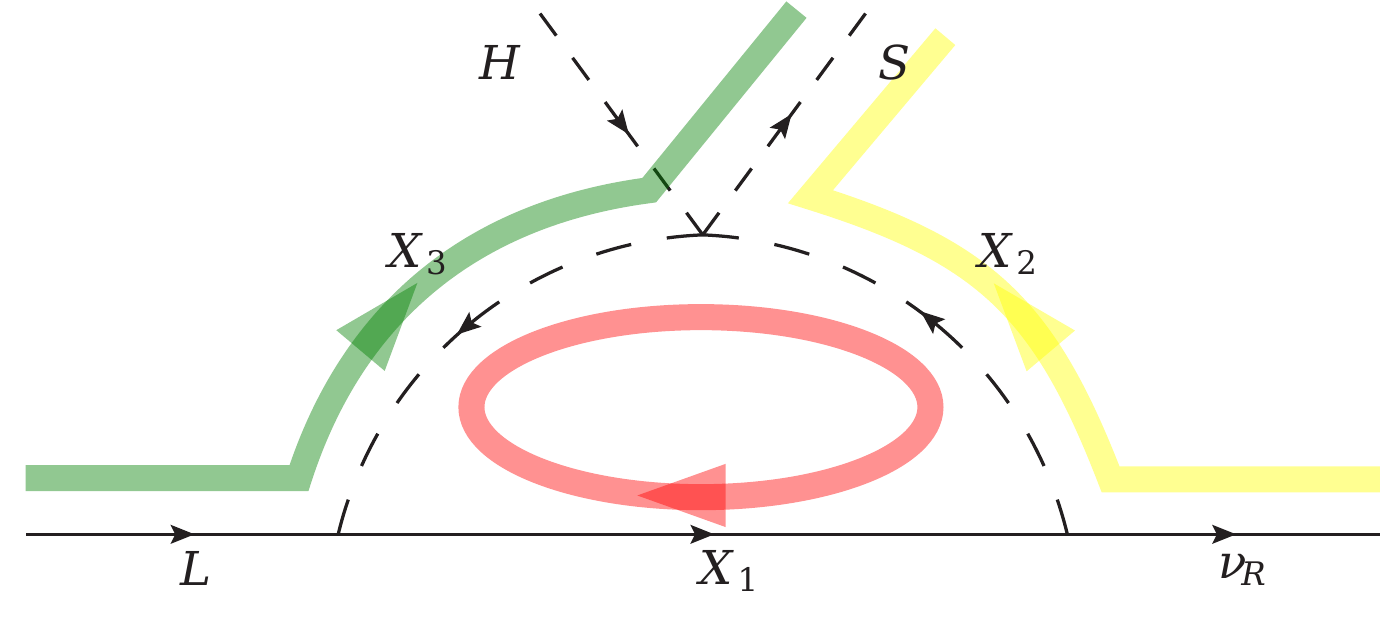}\\
  T3-1-A\\
  \includegraphics[scale=0.55]{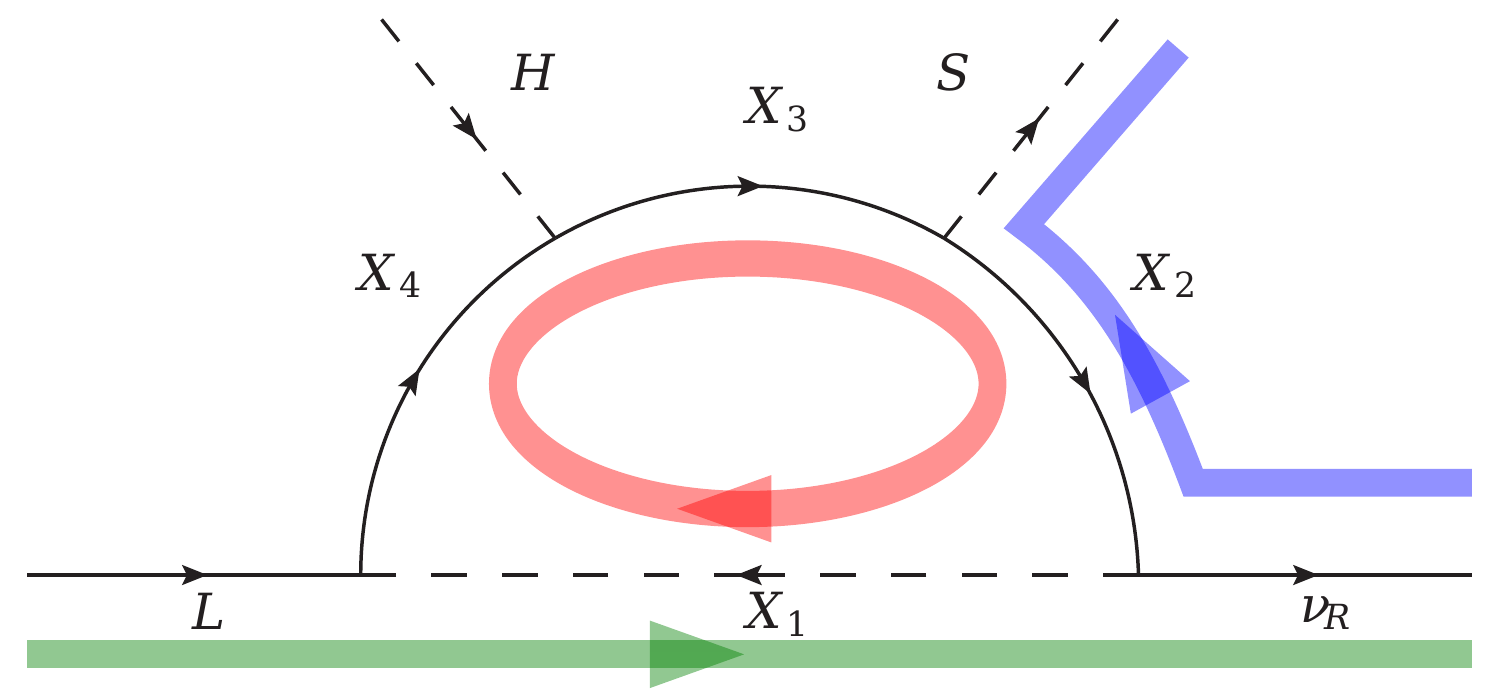}
  \includegraphics[scale=0.55]{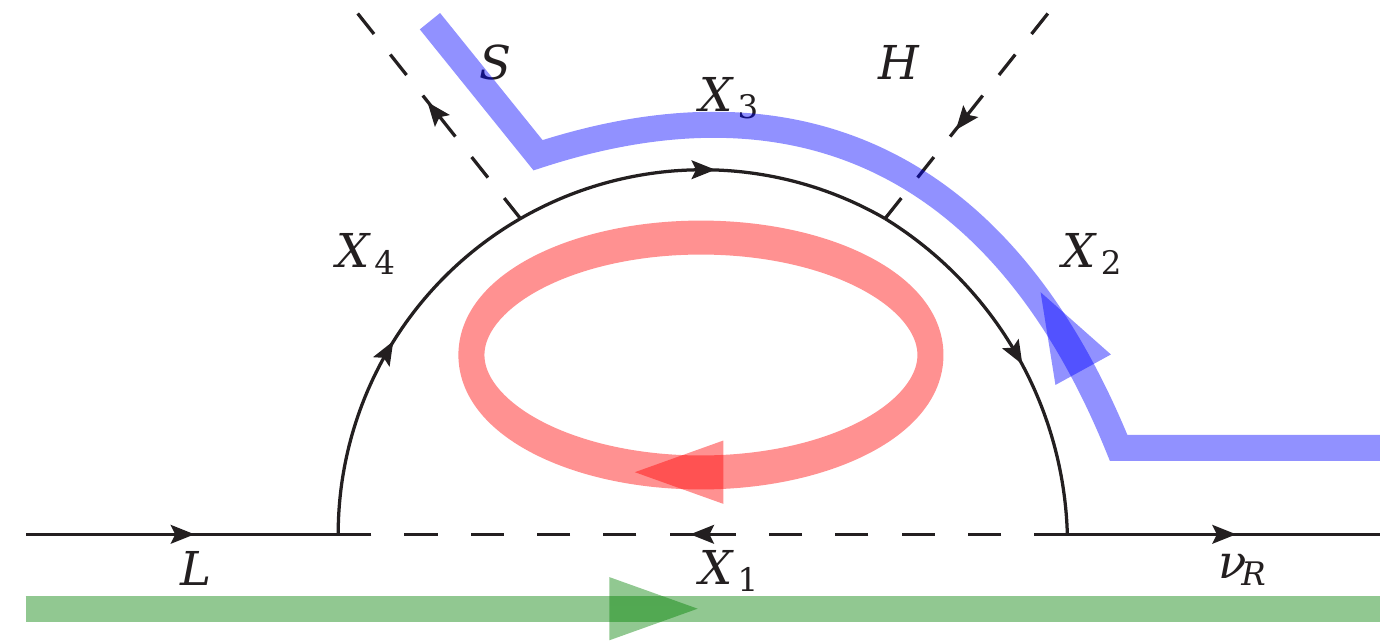}\\
  T1-2-A \hspace{4cm} T1-2-B\\
  \caption{Topologies (in notation of~\cite{Yao:2018ekp}) leading to one-loop Dirac neutrino masses. }
  \label{fig:topo}
\end{figure}

We use the notation for the topologies defined in~\cite{Yao:2018ekp},
which are displayed in figure~\ref{fig:topo}.
There the flux of lepton number is illustrated by the wide colored arrows.
The green arrow represents the flux of the doublet lepton number,
$L(L_i)=-1$, the yellow is for $L(\nu_{R\beta})=-\nu$ with $\beta$
at least $1,2$, the blue for $L(S)=s$, and the red for some
internal circulating $L$ charge associated with a chiral fermion,
which we choose as a free parameter in our setup.

We do not consider the T1-1 or T4 topologies of~\cite{Yao:2018ekp} because the
former is already included in topology T3-1 when the $X_3$
scalar field is decoupled, whereas the latter requires further
symmetries to forbid the tree-level contribution\footnote{
The exception cases are the IV and V solutions of T4-3-I. However, they 
require mixing between the charged leptons and the new fermion
fields, thus leading to charged lepton flavor violation
processes which are quite constrained.}.
In~\cite{Yao:2018ekp} it was assumed that the internal fermion lines were already
vectorlike fermions, which allow them to use the $\nu=1$ solution for
anomaly cancellation conditions, but they needed to impose additional $Z_2$ symmetries to forbid the tree-level contribution to neutrino masses  and to stabilize the dark matter.
Here, we instead assume chiral fields for the fermions that are singlets under
the SM gauge group.  This allow us to search for \emph{new} minimal realizations of the
topologies with a single extra symmetry beyond the SM. 
It is clear that our solution with three chiral states, corresponding
to the three right handed neutrinos, can be easily extrapolated to all
the solutions found in~\cite{Yao:2018ekp}, with the advantage that not
further discrete symmetries are imposed.
We explicitly illustrate this for the case of topology T3-1-A, whose
minimal solution involves just a Dirac fermion.

In our setup, the only extra symmetry beyond the SM, $\operatorname{U}(1)_{B-L}$, must forbid the tree-level terms
\begin{align}
  \mathcal{L}_{\nu}=y^{D}_{\beta i}\left( \nu_{R\beta} \right)^{\dagger}  L_i\cdot  H
  +M^R_{\beta\gamma}\left( \nu_{R\beta } \right)^{\dagger}\left( \nu_{R\gamma} \right)^{\dagger}S+\text{h.c}\,,
\end{align}
as well as allowing for the dimension five operator~\eqref{eq:blop}. 
This is accomplished by the $L$ charge assignments in
Table~\ref{tab:tlcon}. There, $\nu$ is the lepton number of the left-handed antineutrino, which is common to at least the two right-handed neutrinos required to explain the neutrino oscillation data. As already mentioned, the solution with $\nu=1$, studied in~\cite{Yao:2018ekp}, is no longer considered in this work.

\renewcommand{\arraystretch}{2.2}
\begin{table}
  \centering
  \begin{tabularx}{1.0\linewidth}{X|XXXX}\hline
    Fields& $H$ & $L_i$ & $\left( \nu_{R\beta} \right)^{\dagger} $ & $S$ \\\hline
     $L$  &  0  & $-1$  & $\nu\ne 1$                                & $\nu-1\ne -2\nu $\\\hline
  \end{tabularx}
  \caption{General assignment of lepton number for external legs of the one-loop topologies in figure~\ref{fig:topo}}
  \label{tab:tlcon}
\end{table}

To find the new possible solutions, we explore the anomaly
cancellation conditions with five chiral fields by checking that
their charges do not generate direct or induced Majorana mass terms for them, that is, the fermion loop mediators are Dirac-like fields. 
In addition to the chiral fields for $\beta=1,2$, we introduce
$\left( \nu_{Rk} \right)^{\dagger}$ with $L(\nu_{Rk})=-\nu_k$, and a
heavy Dirac fermion field with Weyl components $\psi_L$ and
$\left( \psi_R \right)^{\dagger}$, such that $L(\psi_L)=l$ and
$L(\psi_R)=-r$ respectively.
The linear and cubic anomaly cancellation conditions are~\cite{Batra:2005rh,Nanda:2017bmi}
\begin{align}
  \label{eq:lincub}
  2\nu + \nu_k +l + r=3\,,\quad & 2\nu^3 +\nu_k^3+r^3+l^3=3\,.
\end{align}
From the linear equation
\begin{align}
  \label{eq:nukgen}
  \nu_k=3-2\nu-l-r\,.
\end{align}

To further proceed, we separate the topologies in two types: The
set (A) with T1-3-D and T3-1-A, corresponding to the ones with the
yellow $\nu$ flux in figure~\ref{fig:topo}, and where $S$ is in a
vertex only involving scalars; and the set (B) with T1-3-E and T1-2-(A/B), with the blue $s$
flux in figure~\ref{fig:topo}, and where $S$ is in a Yukawa-type
vertex.
For each case we have:
\begin{itemize}
\item[(A)] The heavy Dirac fermion has a vectorlike mass, such that
  \begin{align}
    l=-r\,.
  \end{align}
  Replacing back in eq.~\eqref{eq:nukgen} we obtain
  \begin{align}
    \nu_k=&3-2\nu\,.
  \end{align}
  Solving the cubic equation for $\nu$ gives rise to
  two different roots: $1$ and $4$.
  Thus, we choose $\nu=4$ which leads to $\nu_k=-5$.

  On the other hand, the fact that the new fermion field $\psi$
  does not contribute
  to the anomalies implies that the only possible realization within
  the T1-3-D topology are the solutions I and II
  with $\psi_L=L_i$ and $\psi_R=e_{Ri}$,
  since in these cases the corresponding contributions
  to the anomalies are already taken into account. 

\item[(B)] The two SM-singlet chiral fields can acquire a Dirac mass (after the spontaneous symmetry breaking (SSB) of $\operatorname{U}(1)_{B-L}$) through
  \begin{align}
    \label{eq:DiracS}
  \mathcal{L}_\psi={h_S} \left( \psi_R \right)^{\dagger} \psi_LS  +\text{h.c}.\,.
\end{align}
If we choose $r$ as the free charge  circulating in the loop, and since from Table~\ref{tab:tlcon} we have $s=\nu-1$, then from the condition in eq.~\eqref{eq:DiracS}: $r+l+s=0$, we get
\begin{align}
  \label{eq:Ll}
  l=&1-\nu-r\,,
\end{align}
and  replacing back in eq.~\eqref{eq:nukgen}
\begin{align}
  \label{eq:nuk5}
  \nu_k=& 2-\nu\,.
\end{align}

Using~\eqref{eq:Ll} and \eqref{eq:nuk5} in the cubic condition for
anomaly cancellation, eq.~\eqref{eq:lincub}, we end up with the
one-parameter solution
\begin{align}
  \label{eq:five}
  \nu=\frac{r^2-r+2}{3-r}\,.
\end{align}
If $\nu_k=0$ we would have a solution with four chiral fields when
$\nu=2$, however the required $r$ charge is irrational and will not be
further considered here.
\end{itemize}

To label the solutions we use the conventions of \cite{Yao:2018ekp}, as
Ta-n-b-I~$\alpha$, where ``a'' refers to the topology itself, ``n''
indicates the different choices of the fermion and scalar lines in a
given topology, ``b'' denotes the field assignments for external
fields, ``I'' denotes the simplest solution with standard model
singlet scalars or fermions, and $\alpha$ is the parameter which
fixes the hypercharges of the $X_i$ fields inside the loop.

It is worth mentioning that  when both $\psi_L$ and $\psi_R$ are SM singlets
(in the T3-1-A-I with $\alpha=0$, T1-3-E-I with $\alpha=0$,
T1-2-A-I with $\alpha=0$ and T1-2-B-II with $\alpha=+1$ models), 
the condition $r\ne 1$ or $l\ne -1$
must be imposed in order to avoid the tree-level realization of~\eqref{eq:blop}
that involves a SM singlet fermion mediator --the so-called type I Dirac seesaw.
In this way one of the two required terms in that realization 
($\psi_R^{\dagger} L_i\cdot H$ and $ \nu_{R\beta}^{\dagger} \psi_L S^*$)
is forbidden. 


As usual for scotogenic models, we demand  the lightest neutral particle running in the loop to be stable.
For the case of scalar DM, the stability is guaranteed if there is no linear in the scalar loop mediators in the scalar potential neither Yukawa interactions with two SM fermions.
The minimal DM scenarios that may arise are then the singlet~\cite{Silveira:1985rk,McDonald:1993ex,Burgess:2000yq}, doublet~\cite{Deshpande:1977rw,Barbieri:2006dq} and singlet-doublet~\cite{Kadastik:2009dj,Kadastik:2009cu,Kakizaki:2016dza,Liu:2017gfg} scalar DM.
It is worth mentioning that since Majorana mass terms for the fermion loop mediators are not allowed, the fermion DM candidate is 
 either singlet~\cite{Kim:2006af} or singlet-doublet Dirac DM~\cite{Yaguna:2015mva}.

The wanted solutions must satisfy the constraints regarding DM stability and Diracness of light neutrinos, and guarantee that the direct or induced fermion mass terms between the right handed neutrinos and $\psi_L$ or $\psi_R$  are forbidden.
  The reason to exclude this kind of mixings is that in such a case the DM would not be stable,  because the fermion loop mediator would decay into particles of the visible sector.
 For instance, the induced (through $S$) or direct mixing between $(\nu_{Rk})^\dagger$ with either $\psi_L$ or $(\psi_R)^\dagger$ leads to the decay into $S$ and $\nu_{Rk}$ for the induced mixing, and into $Z_\mu'$ and $\nu_{Rk}$ for the direct mixing.

The solutions of the two sets are displayed in Table~\ref{tab:sltns}. Solution (A) is the well known one studied in~\cite{Ma:2014qra} for
tree-level realization of the 5-dimensional operator.
In this solution $r$ is quite free, in fact $r=\pm 1/2,\pm 1/3,\ldots$
The solution (B) was obtained after exploring all the solutions of
eq.~\eqref{eq:five} for rational values of $|r|\le 10$, and with both
the numerator and denominator less or equal than $10$.
Since the anomaly cancellation conditions are invariant under the exchange of $r$ with $l$, a second solution for (B) exists with the charges of $\psi_L$ and $(\psi_R)^\dagger$ exchanged.

\begin{table}
\centering
\def\nc{2}
\renewcommand{\arraystretch}{2.2}
\begin{tabularx}{\textwidth}{XX|XXXXXX}\hline
 Fields  & & $\left( \nu_{Ri} \right)^{\dagger}$
  &$\left( \nu_{Rj} \right)^{\dagger}$&$\left( \nu_{Rk} \right)^{\dagger}$&$\psi_L$&$\left( \psi_R \right)^{\dagger}$&$S$\\ \hline
  \multirow{2}{*}{$L$}&(A)&$+4$&$+4$&$-5$& $-r$ & $r$ & $+3$\\
&(B) &$\displaystyle{+\frac{8}{5}}$&$\displaystyle{+\frac{8}{5}}$&$\displaystyle{+\frac{2}{5}}$&$\displaystyle{\frac{7}{5}}$&$\displaystyle{-\frac{10}{5}}$&$\displaystyle{+\frac{3}{5}}$\\\hline
\end{tabularx}
\caption{Solutions for Dirac neutrino masses with Dirac loop mediators for $i\ne j\ne k$.  }
\label{tab:sltns}
\end{table}


Higher $\operatorname{SU}(2)_L$ fermion representations, as required in
T1-2 topologies, need to be introduced as vectorlike fermions to not
spoil the anomaly cancellation conditions of the standard model.
We will denote vectorlike doublet Weyl fermions fields with $Y=- 1$
($Y=+1$) as $\Psi_{L,R}$ ($\Upsilon_{L,R}$).

Regarding the scalars circulating in the loop, we will use $\sigma$ and
$\eta$ to represent $\operatorname{SU}(2)_L$ scalar singlets and
scalar doublets respectively, and we will denote their non-zero lepton
number with the same symbols.

A final comment is in order. Because  $\operatorname{U}(1)_{B-L}$ is promoted to a gauge symmetry, the vacuum expectation value of $S$, $\langle S\rangle=v_S/\sqrt{2}$, induces a non-zero mass to the associated gauge boson $Z_{BL}$.
  The expression for its mass can be cast as $M_{Z_{BL}}=g_{BL}v_S|s|$, where $g_{BL}$ is the $B-L$ gauge coupling  and $s$ is the $B-L$ charge of $S$.
  On the other hand, since $Z_{BL}$ couples to all the SM fermions (they have non zero $B-L$ charges) it can be produced in hadron and lepton colliders leading to observable signatures. Indeed, from the non-observation of any of such signatures in the LEP and LHC data there exist constraints on its mass and gauge coupling~\cite{Carena:2004xs,Cacciapaglia:2006pk,Aaboud:2017buh,Escudero:2018fwn,Sirunyan:2018exx} (see e.g. \cite{Wang:2017mcy,Han:2018zcn} for specific analysis in $B-L$ scotogenic Dirac models). 

\section{Solutions}
\label{sec:sol}

The new solutions correspond to the case in which some of the $X_i$ fermion
fields in figure~\ref{fig:topo} can be chosen as chiral fields.
We explore the solutions with the minimal number of fermion fields beyond
the stadard model. We are interested, therefore, in the solutions in which at least two
right handed neutrinos have the same $\operatorname{U}(1)_{B-L}$
charge, because in such a case both of them can couple to the same set
of extra chiral fermions.
All the solutions for the $B-L$ charges presented below have been
choosen in such a way the DM particle does not decay.

\subsection{Chiral T1-3-D-I ($\alpha=-2$)}

We will start our analysis with the case in which an internal fermion
line in figure~\ref{fig:topo} can be interpreted as a standard
model field. Therefore, only the right handed neutrinos contribute to the anomaly
cancellation conditions of the SM with $\operatorname{U}(1)_{B-L}$.
There are three well known solutions with three chiral
fields~\cite{Nomura:2017jxb,Ma:2014qra}.
However, solution (A) in Table~\ref{tab:sltns}, is the only one that
satisfies our constraints.

\begin{figure}
\centering
\includegraphics[scale=0.55]{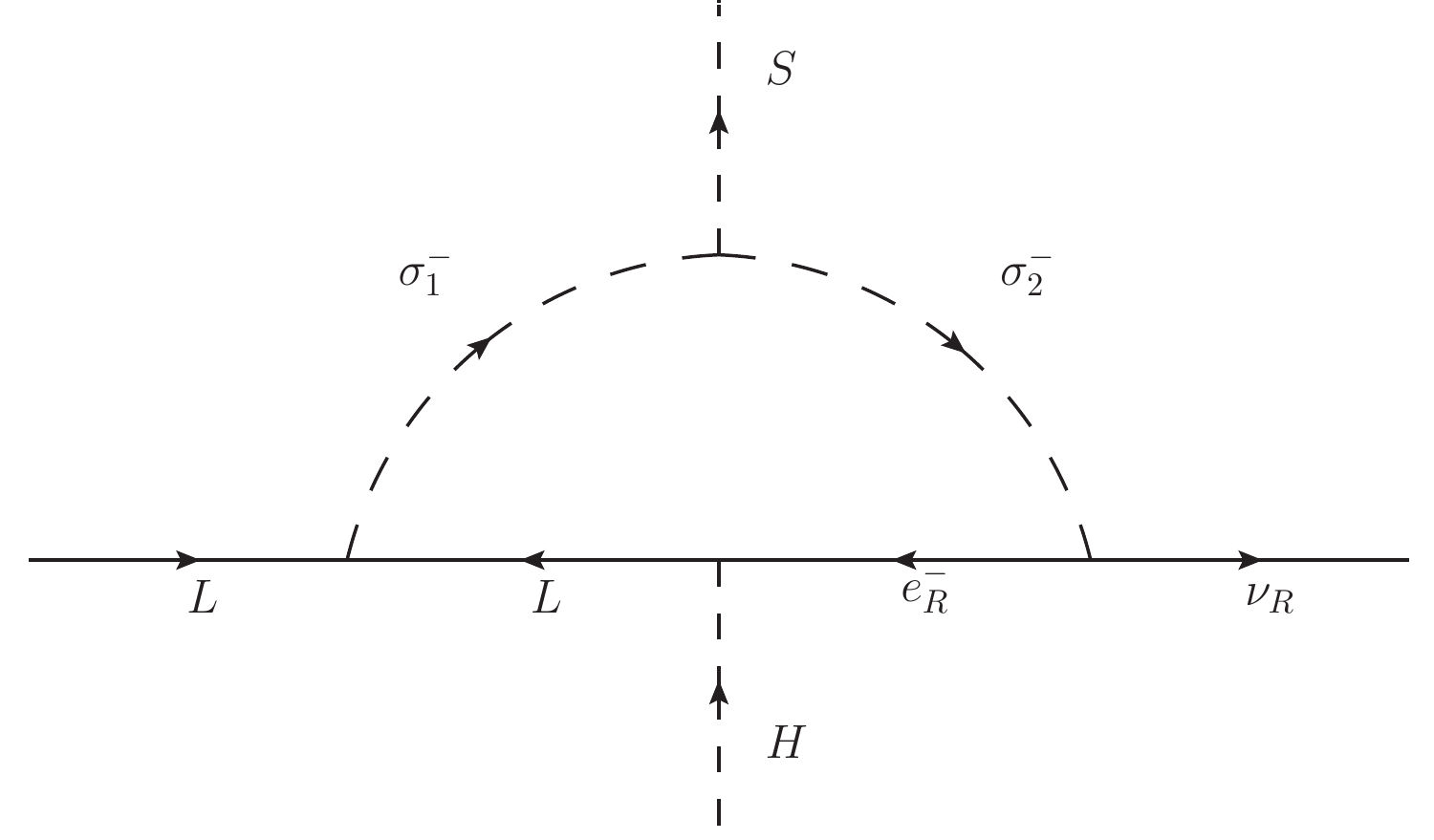}

Chiral T1-3-D-I ($\alpha=-2$) 
\caption{Dirac Zee model}
\label{fig:zee}
\end{figure}

In fact, with the additions of two charged scalars,
$\sigma_{1,2}^{\pm}$, which are singlet under $\operatorname{SU}(2)_L$,
we can build the Dirac version of the Zee mechanism to generate
neutrino masses~\cite{Kanemura:2011jj}.
The couplings required to build the diagram displayed in
figure~\ref{fig:zee} are
\begin{align}
  \mathcal{L}\subset \left[ f_{ij}L_i\cdot L_j\, \sigma_1^{+}
  + h_e^{ij}\left( e_{Ri} \right)^{\dagger} L_j\cdot \widetilde{H}
  +h_R^{i\beta}\,e_{Ri}\,\nu_{R\beta}\sigma_2^++\text{h.c} \right]
  +V(\sigma_1^{\pm},\sigma_2^{\pm},H)
\end{align}
where $L_i$ are the SM lepton doublets, $H=(H^{+}, H^0)^T$, $\widetilde{H}=i\sigma_2 H^*$
and $V(\sigma_1^{\pm},\sigma_2^{\pm},H)$ is the scalar potential.

Therefore, $\psi_L$ ($\psi_R$) in solution (A) of
Table~\ref{tab:sltns} corresponds to three lepton doblets $L_i$
(right-handed electrons $e_{Ri}^{-}$) with $l=-1$ ($r=1$) as the usual
lepton number.  Since we use one set of charged scalar fields,
$\sigma_{1,2}^{\pm}$, the model has two massless chiral fields, one of
them,
$\nu_{Rk}$, contributing to effective number of relativistic degrees
of freedom,
$N_{\text{eff}}$~\cite{Nomura:2017jxb,Perez:2017qns,Han:2018zcn}.
%
From the Lepton number flux in figure~\ref{fig:zee}, we have 
\begin{align}
  L(\sigma_1^+)=&-2\,,&   L(\sigma_2^+)=&-5\,.
\end{align}
The full solution is presented  in Table~\ref{tab:zee}. This is by far the minimal model for Dirac neutrino masses with a gauged $\operatorname{U}(1)_{B-L}$. The Dirac Zee model with $\nu=\nu_k=1$ and extra discrete symmetries has been studied in~\cite{Kanemura:2011jj}.

\renewcommand{\arraystretch}{1.8}
\begin{table}
\centering
\def\nc{2}
\begin{tabularx}{\textwidth}{X|XXXXXXXXXXX}\hline
Fields & \multirow{\nc}{*}{$\left( \nu_{Ri} \right)^{\dagger}$}
 &\multirow{\nc}{*}{$\left( \nu_{Rj} \right)^{\dagger}$}&\multirow{\nc}{*}{$\left( \nu_{Rk} \right)^{\dagger}$}&\multicolumn{2}{Y}{$X_1$}&\multicolumn{2}{Y}{$X_2$}&$X_3$&\multicolumn{2}{Y}{$X_4$}&\multirow{\nc}{*}{$S$}\\  
$\text{T1-3-D-I}$&&&&\multicolumn{2}{Y}{$e_{R_l}^-$}&\multicolumn{2}{Y}{$\sigma_2^-$}&$\sigma_1^-$&\multicolumn{2}{Y}{$L_l$}&\\\hline
$L$&$+4$&$+4$&$-5$&\multicolumn{2}{Y}{$-1$}&\multicolumn{2}{Y}{$-5$}&$-2$&\multicolumn{2}{Y}{$-1$}&3\\\hline
\end{tabularx}
\caption{Chiral T1-3-D-I ($\alpha=-2$): Solutions for the Dirac Zee model with $i\ne j\ne k$ ($i,j,k,l=1,2,3$).}
\label{tab:zee}
\end{table}

It is worth noticing that the restrictions from $N_{\text{eff}}$ are
expected to be stronger in our model because of the larger lepton
number assignment for the right-handed neutrinos. However, since they do not couple directly to any SM particles, we can simply assume that their interaction with the extra gauge boson and scalars are sufficiently suppressed that they  decouple early enough from the thermal bath. 
On the other hand, it is clear that there is not a DM candidate in this model. Indeed, this is just a specific example of models with one-loop Dirac neutrino masses but without a DM candidate that can be obtained within our setup.

Regarding $\nu_{Rk}$, it can give rise to either a third Dirac
neutrino mass if we extend the scalar sector with $S'$ and
${\sigma'}_2^{\pm}$ of $L$ charges $-6$ and $+8$ respectively, or a Majorana dark
matter candidate if we extend the scalar sector with a $S'$ of $L$
charge $+10$~\cite{Nomura:2017jxb}.  In both cases we end up with a
physical Goldstone boson (GB) which could contribute to
$N_{\text{eff}}$ through interactions with the
Higgs~\cite{Weinberg:2013kea}.
The conditions that GB decouples from the bath in the early universe
are analyzed in~\cite{Nomura:2017jxb} and require couplings of GB with
SM Higgs not larger than $10^{-3}$.
This discussion can be easily extended to the other one-loop
realizations below.

\begin{figure}
\centering
\includegraphics[scale=0.6]{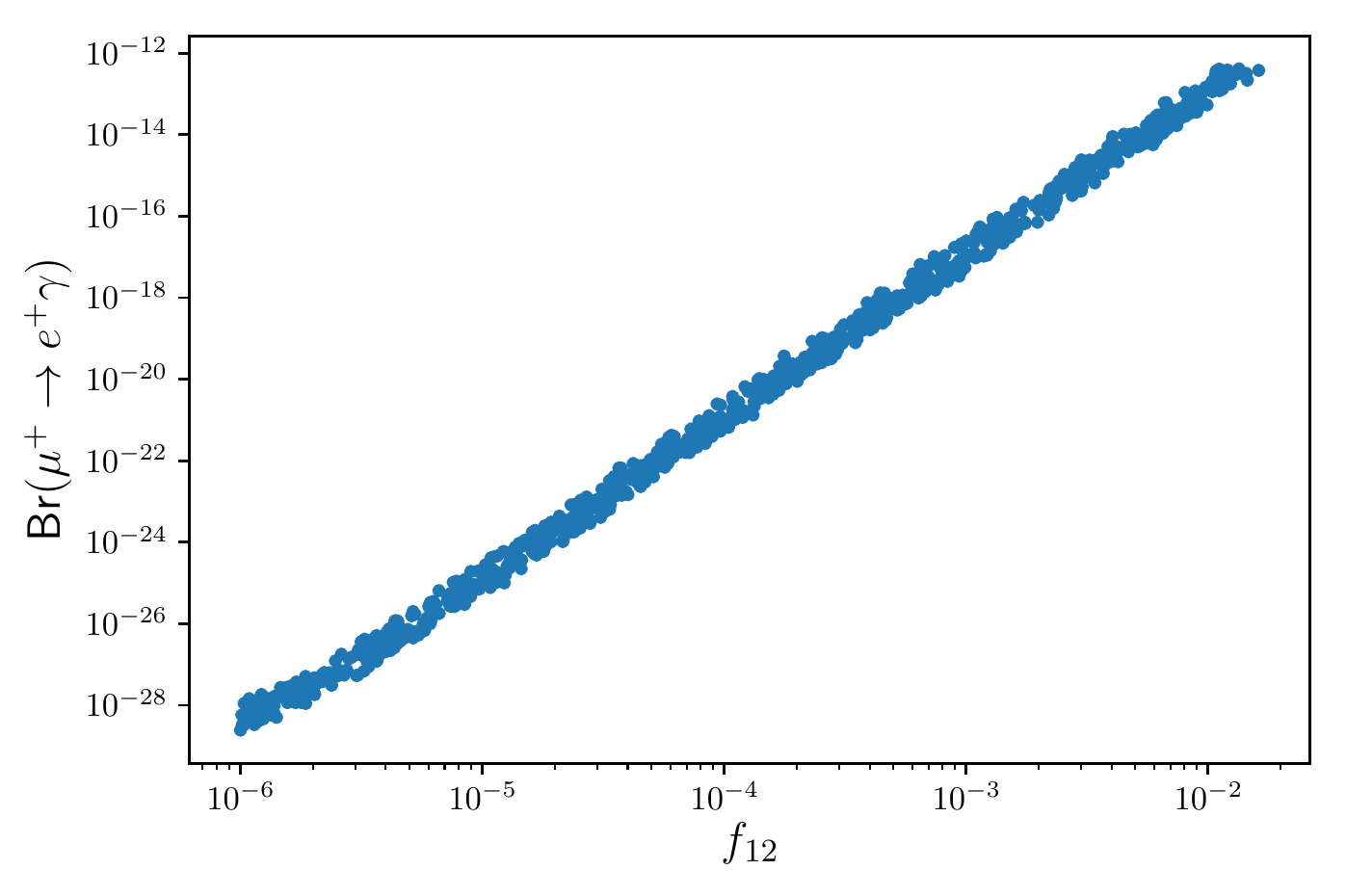}
\caption{Relationship between $\text{Br}(\mu^{+} \to e^{+} \gamma)$ and the parameter $f_ {12}$}
\label{fig:muegamma}
\end{figure}

We have implemented the model with three non-zero Dirac neutrino
masses in SARAH~\cite{Staub:2013tta}.
We use the method in~\cite{Kanemura:2011jj,Ludl:2014axa} to express $f_{13}$, $f_{23}$, $h_{11}$, $h_{22}$, $h_{33}^R$, $h_{31}^R$, $h_{32}^R$ and $h_{23}^R$ as a function of the neutrino masses and mixings, by
using  $f_{12}^R$, $h_{12}^R$, $h_{13}^R$, $h_{21}^R$ as free parameters.
As an example of the consistency of the model, we show in
figure~\ref{fig:muegamma} the observable
$\text{Br}(\mu^{+} \to e^{+} \gamma) < 4.2 \times 10^{-13}$~\cite{TheMEG:2016wtm} as function a $f_{12}$.
The other parameters were fixed as $\theta = 0.1$, $M_{\sigma_1} = 500\ \text{GeV}$, $M_{\sigma_2} = 750\ \text{GeV}$, $h_{12}^R$, $h_{13}^R$, $h_{21}^R = 10^{-4}$, where $\theta$ is the mixing angle between the charged mass eigenstates $\sigma_1$ and $\sigma_2$.
We can see that the value for the parameter $ f_ {12} $ is restricted
to values lower than 0.02.

\subsection{Chiral T1-3-E-I ($\alpha=0$)}
\label{sec:chiral-t1-3}
We consider now the topologies where two SM-singlet chiral
fields acquire a Dirac mass after the SSB of
$\operatorname{U}(1)_{B-L}$ from the term $\mathcal{L}_{\psi}$ in eq.~\eqref{eq:DiracS}. 
For them we will use the solution (B). 
The relevant terms in the Lagrangian include
\begin{align}
  \mathcal{L}\supset\mathcal{L}_{\psi} +\mathcal{L}_{\sigma\psi}+\mathcal{L}_{\eta\psi} +V(\sigma_a,\eta_a,S,H),
\end{align}
where $\mathcal{L}_{\psi}$ was given in eq.~\eqref{eq:DiracS}, 
\begin{align}
  \mathcal{L}_{\sigma\psi}=&h_1^{\beta a} \left( \nu_{R\beta} \right)^{\dagger} \psi_L\, \sigma^{*}_a+\text{h.c}\nonumber\\
  \mathcal{L}_{\eta\psi} =&{h'}_1^{i a} \left( \psi_R \right)^{\dagger} L_i\cdot \eta_a +\text{h.c}\,,
\end{align}
and $V(\sigma_a,\eta_a,S,H)$ is the scalar potential.

After the spontaneous breaking of the
$\operatorname{U}(1)_{B-L}$ symmetry, this topology is reduced to the well
known Dirac radiative seesaw model, but with different Lepton number
assignments.
In fact, the model with $\nu=\nu_k=r=1$ and two extra $Z_2$ discrete
symmetries was first introduced in~\cite{Gu:2007ug,Farzan:2012sa}, while the case with
$r\ne1$ which requires only one extra $Z_2$ symmetry was studied
in~\cite{Reig:2018mdk}.
The minimal set of fermion fields is achieved when the two non-zero
Dirac neutrino masses are generated with two set of SM singlet and
doublet scalars: $\eta_{a}\,,\sigma_a$~\cite{Reig:2018mdk}.

\begin{figure}
\centering
\includegraphics[scale=0.5]{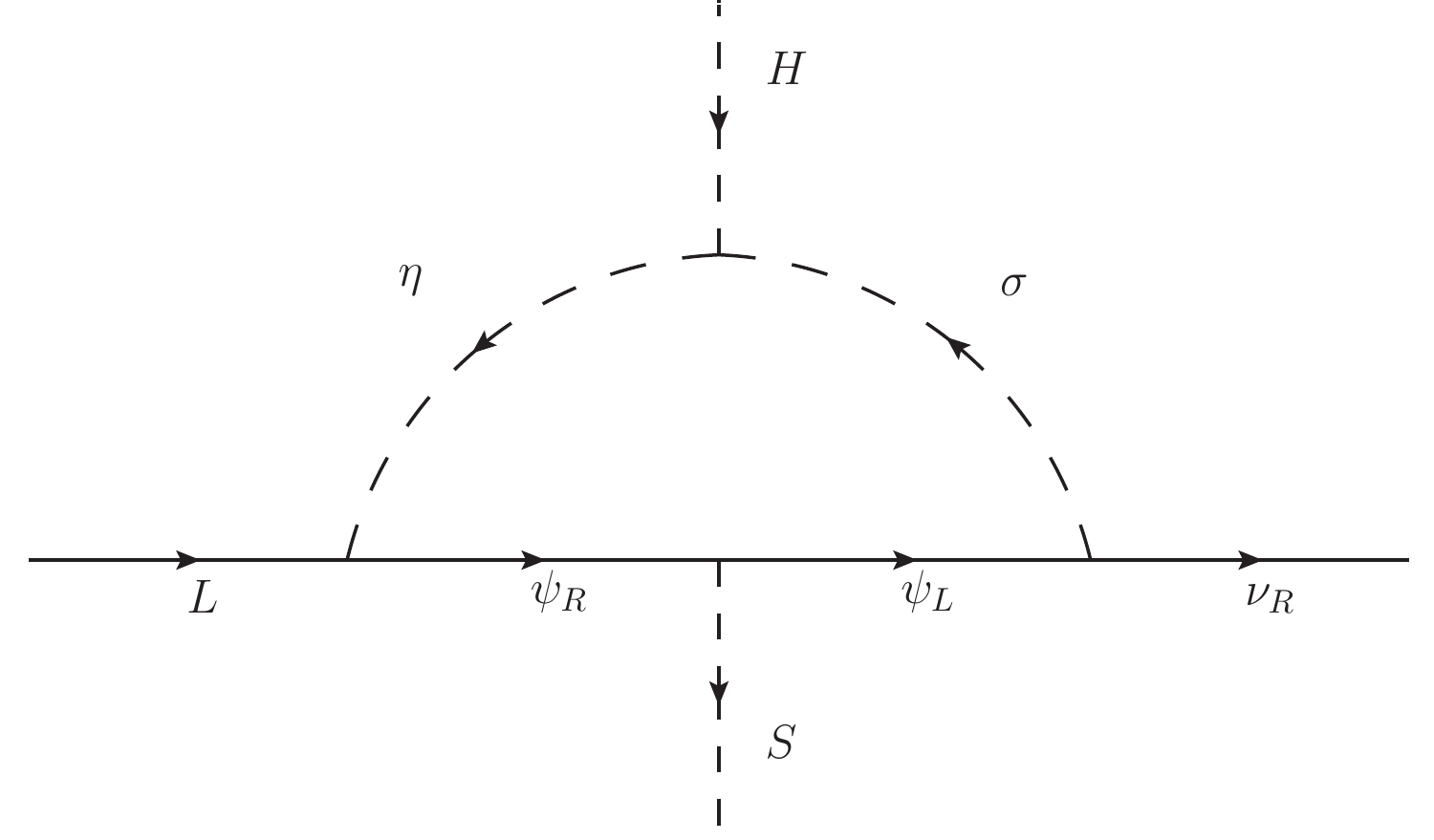}
\caption{T1-3-E-I ($\alpha=0$)}
\label{fig:t13e1}
\end{figure}

From the figure~\ref{fig:t13e1}, the charges of the scalars are 
\begin{align}
  \eta&=1-r\,, &  \sigma=1-r \,.
\end{align}
The solution compatible with the fields in figure~\ref{fig:t13e1} is
displayed in Table~\ref{tab:t1-2-a}.
The phenomenology of the radiative seesaw model with $\nu=1$ has been already studied in the literature~\cite{Gu:2007ug,Farzan:2012sa,Wang:2017mcy,Reig:2018mdk,Han:2018zcn}, where either singlet Dirac ($\psi$) or singlet-doublet scalar ($\sigma^a,\eta^a$) DM is realized.
Because of the similar charges associated to solution (B), we do not expect significant  differences with respect to those works.

\subsection{Chiral T1-2-A-I $\alpha=0$}

\begin{figure}
\centering
\includegraphics[scale=0.5]{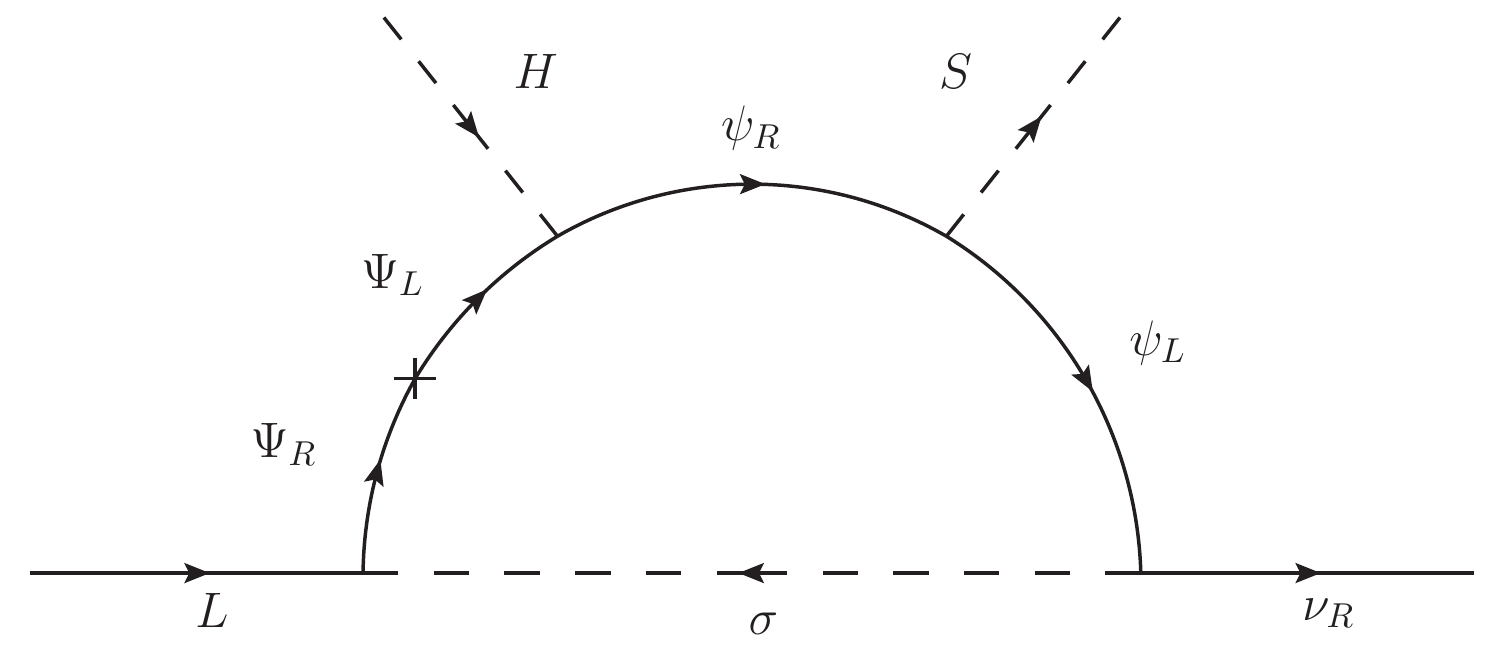}
\includegraphics[scale=0.5]{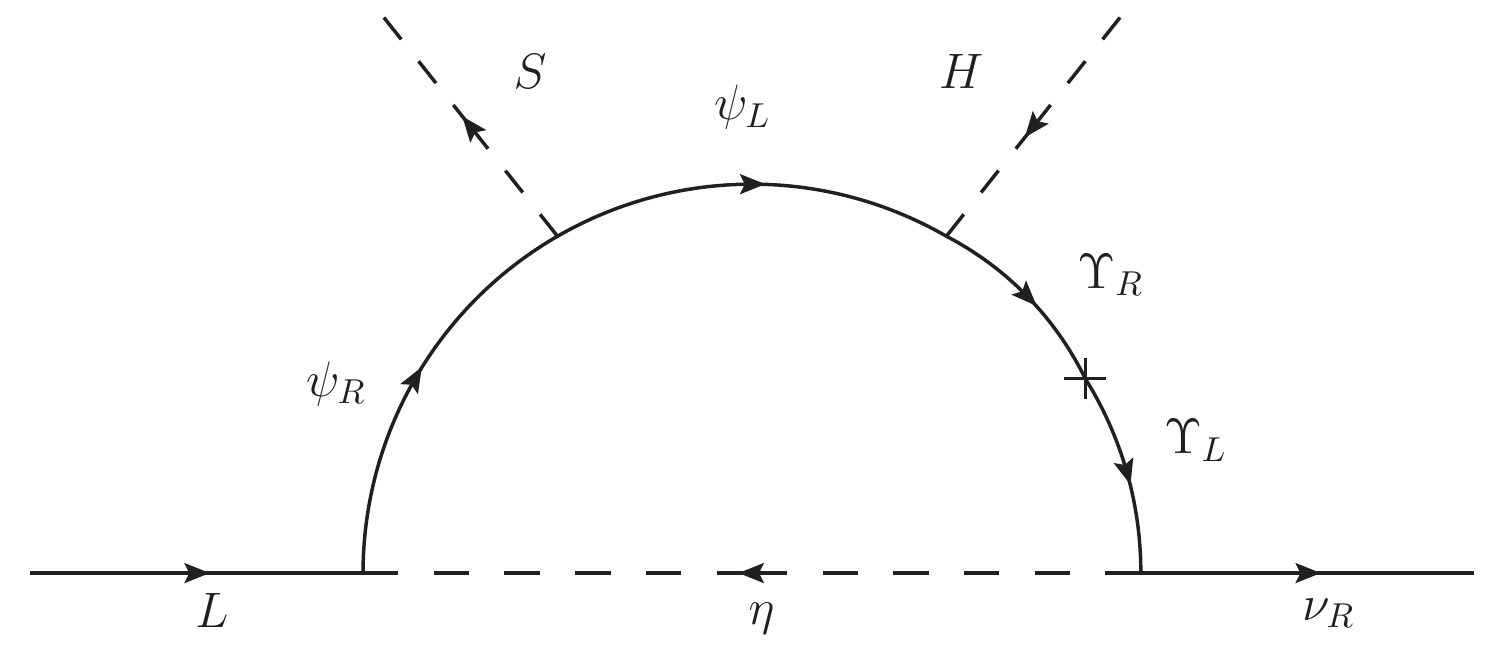}
\caption{Chiral T1-2-A-I ($\alpha=0$)\hspace{2.5cm} Chiral T1-2-B-II ($\alpha=0$)}
\label{fig:t1-2-ab}
\end{figure}

The solution compatible with the fields in figure~\ref{fig:t1-2-ab}
(left), requires at least the following terms in the Lagrangian
\begin{align}
  \mathcal{L}\supset&
  \mathcal{L}_{\psi}+\mathcal{L}_{\sigma\psi}
+ \left[M_{\Psi}\widetilde{\left( \Psi_R \right)}\cdot  \Psi_L
  +h_2^{ia} \widetilde{\left( \Psi_R \right)}\cdot L_i\, \sigma_a  + y_1 \left( \psi_R \right)^{\dagger} \Psi_L \cdot H +\text{h.c}  \right]
+V(\sigma_a,S,H)\,,
\end{align}
where $\Psi_L=\left( \Psi_L^0,\, \Psi_L^-\right)^{\operatorname{T}}$, $\widetilde{\left( \Psi_R \right)}=\left( (\Psi_R^-)^\dagger,\, -(\Psi_R^0)^\dagger\right)^{\operatorname{T}}$
and $V(\sigma_{a},S,H)$ is the scalar potential.
It follows that this model requires a set of at least two SM-singlet scalars to generate a rank-2 neutrino mass matrix, and allows for either singlet scalar ($\sigma^a$) or singlet-doublet Dirac ($\psi, \Psi$) DM.

From the figure~\ref{fig:t1-2-ab} (left)
\begin{align}
  \sigma=1-r\,.
\end{align}

The corresponding charges for the solution (B) are shown in
Table~\ref{tab:t1-2-a}.

\subsection{Chiral T1-2-B-II $\alpha=0$}
The solution compatible with the fields in figure~\ref{fig:t1-2-ab} (right), requires at least the following terms in the Lagrangian
\begin{align}
  \mathcal{L}\supset&
  \mathcal{L}_{\psi}
  +\mathcal{L}_{\eta\psi}
   +\left[ M_{\Upsilon}\widetilde{\left( \Upsilon_R \right)}\cdot  \Upsilon_L
  +{h'}_2^{a\beta} \widetilde{\left( \Upsilon_L \right)}\cdot \eta_{a}\, \nu_{R\beta} + {y'}_1\widetilde{\Upsilon_R}  \cdot H\, \psi_L+\text{h.c}  \right] 
+V(\eta_a,S,H)\,,
\end{align}
where $\Upsilon_L=\left(\Upsilon_L^+,\, \Upsilon_L^0\right)^{\operatorname{T}}$, $\widetilde{\left( \Upsilon_R \right)}=\left( (\Upsilon_R^0)^\dagger,\, -(\Upsilon_R^+)^\dagger\right)^{\operatorname{T}}$
and $V(\eta_{a},S,H)$ is the scalar potential.
It follows that this model requires a set of at least two scalar doublets to generate a rank-2 neutrino mass matrix, and allows for either doublet scalar ($\eta^a$) or singlet-doublet Dirac ($\Upsilon, \psi$) DM.

From the figure
\begin{align}
    \eta=1-r\,.
\end{align}
The solutions compatible with the fields in figure~\ref{fig:t1-2-ab}
(right) correspond to the ones displayed in Table~\ref{tab:t1-2-a}.

\renewcommand{\arraystretch}{2.2}
\begin{table}
\centering
\def\nc{2}
\begin{tabularx}{\textwidth}{X|XXXXXXXXXX}\hline
Fields& \multirow{\nc}{*}{$\left( \nu_{Ri} \right)^{\dagger}$}
 &\multirow{\nc}{*}{$\left( \nu_{Rj} \right)^{\dagger}$}&\multirow{\nc}{*}{$\left( \nu_{Rk} \right)^{\dagger}$}&$X_1$&\multicolumn{2}{Y}{$X_2$}&$X_3$&\multicolumn{2}{Y}{$X_4$}&\multirow{\nc}{*}{$S$}\\  
$\footnotesize\text{T1-3-E-I}$&&&&$ \psi_L$&$\eta_a$&&$\sigma_a$&$ \psi_R$&&\\\hline
$L$&\multirow{5}{*}{$\displaystyle{\frac{8}{5}}$}&\multirow{5}{*}{$\displaystyle{\frac{8}{5}}$}&\multirow{5}{*}{$\displaystyle{\frac{2}{5}}$}&$\displaystyle{\frac{7}{5}}$&$\displaystyle{\frac{15}{5}}$&&$\displaystyle{\frac{15}{5}}$&$\displaystyle{\frac{10}{5}}$&&\multirow{5}{*}{$\displaystyle{\frac{3}{5}}$}\\\cline{5-10}
$\footnotesize\text{T1-2-A-I}$&&&&$\sigma_a$&$ \psi_R $&&$\psi_L$&$\widetilde{\left( \Psi_R \right)}$&$\Psi_L$&\\\cline{5-10}
$L$&&&&$\displaystyle{\frac{15}{5}}$&$\displaystyle{\frac{10}{5}}$&&$\displaystyle{\frac{7}{5}}$&$\displaystyle{-\frac{10}{5}}$&$\displaystyle{\frac{10}{5}}$&\\\cline{5-10}
$\footnotesize\text{T1-2-B-II}$&&&&$\eta_a$&$\widetilde{\left( \Upsilon_R \right)}$&$\Upsilon_L$&$ \psi_R $&$\psi_L$&&\\\cline{5-10}
$L$&&&&$\displaystyle{\frac{15}{5}}$&$\displaystyle{-\frac{7}{5}}$&$\displaystyle{\frac{7}{5}}$&$\displaystyle{\frac{10}{5}}$&$\displaystyle{\frac{7}{5}}$&&\\\hline
\end{tabularx}
\caption{Chiral solutions for Dirac neutrino masses with the minimal T1 topologies with $i\ne j\ne k$ and $\alpha=0$.}
\label{tab:t1-2-a}
\end{table}

The phenomenological analysis of the chiral realizations of topologies  T1-2 will be done elsewhere. It is worth noting, however, that a model independent analysis of the effect of the right handed neutrinos  on $N_{\text{eff}}$, can be inferred from~\cite{Perez:2017qns}. The ratio $M_{Z_{BL}}/(\nu g_{BL})$, where $g_{BL}$ is the $\operatorname{U}(1)_{B-L}$ gauge coupling, must be larger than $7$--$8\ \text{TeV}$  in order to be in agreement with the cosmological constraints.

\subsection{vectorlike solutions}
\label{sec:vectorlike-solutions}
In~\cite{Yao:2018ekp}, all the obtained solutions for the topologies
in figure~\ref{fig:topo} have internal vectorlike fermions.
We want to stress that it is possible to realize all of them 
with a single extra symmetry $\operatorname{U}(1)_{B-L}$.
For that,  we can use the solution (A) with a proper choice of the
circulating free charge $r$.

\begin{figure}
\centering
\includegraphics[scale=0.6]{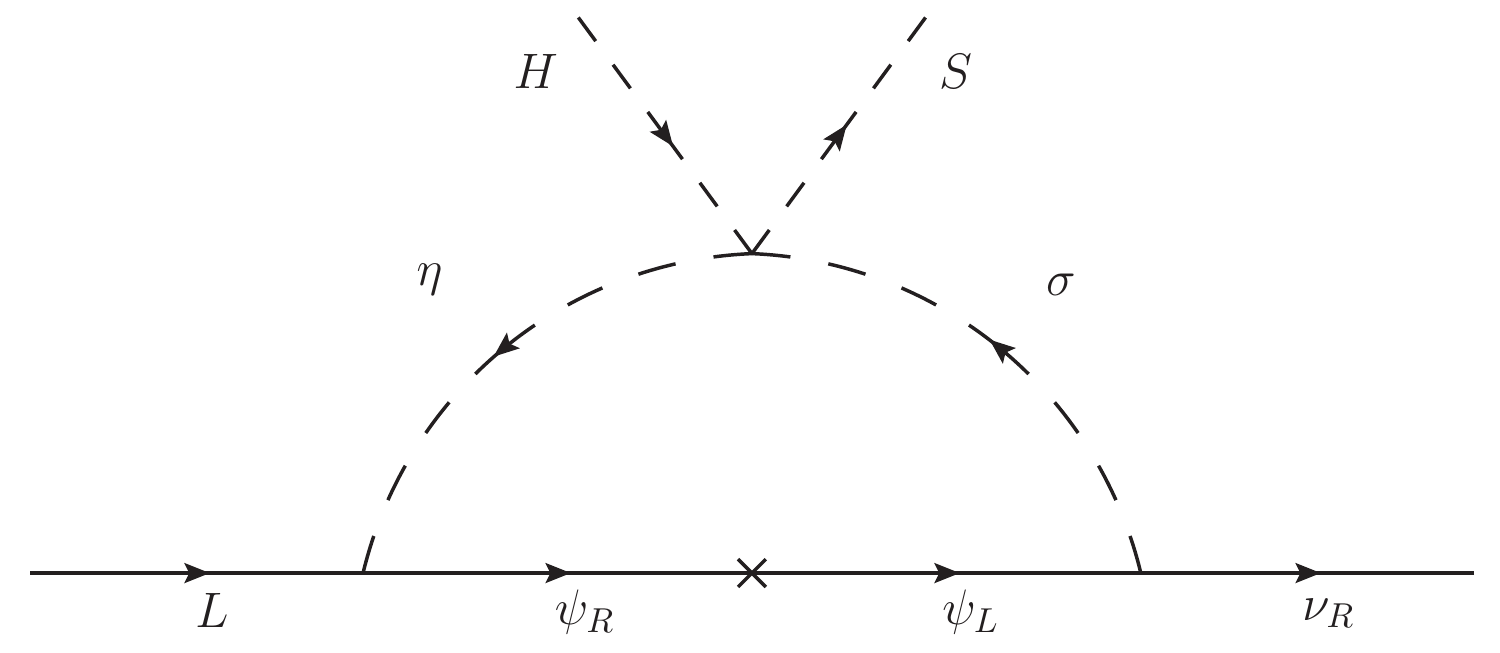}
\caption{T3-1-A-I  ($\alpha=0$): $B-L$ flux in the Dirac  radiative seesaw}
\label{fig:dm0epsilon}
\end{figure}

For example, in the simplest case of just one circulating SM-singlet
Dirac fermion line as in T3-1-A, we have the diagram displayed in
figure~\ref{fig:dm0epsilon}.

From the flux of lepton number in figure \ref{fig:dm0epsilon}, we have. 
\begin{align}
  \eta=&1-r\,,&   \sigma=&\nu-r\,.
\end{align}
The corresponding charges for solution (A) are shown in Table~\ref{tab:t31i}.

\renewcommand{\arraystretch}{2.2}
\begin{table}
\centering
\def\nc{2}
\begin{tabularx}{\textwidth}{X|XXXXXXXXXXX}\hline
Fields& \multirow{\nc}{*}{$\left( \nu_{Ri} \right)^{\dagger}$}
 &\multirow{\nc}{*}{$\left( \nu_{Rj} \right)^{\dagger}$}&\multirow{\nc}{*}{$\left( \nu_{Rk} \right)^{\dagger}$}&\multicolumn{2}{Y}{$X_1$}&\multicolumn{2}{Y}{$X_2$}&$X_3$&\multicolumn{2}{Y}{$X_4$}&\multirow{\nc}{*}{$S$}\\  
$\small\text{T3-1-A-I}$&&&&$\left( \psi_R \right)^{\dagger}$&$\psi_L$&\multicolumn{2}{Y}{$\eta_a$}&$\sigma_a$&\multicolumn{2}{Y}{$-$}&\\\hline
$L$&$+4$&$+4$&$-5$&$r$&$-r$&\multicolumn{2}{Y}{$1-r$}&$4-r$&\multicolumn{2}{Y}{$-$}&3\\\hline
\end{tabularx}
\caption{T3-1-A-I ($\alpha=0$): Solutions for Dirac radiative seesaw model with $i\ne j\ne k$ ($i,j,k=1,2,3$). }
\label{tab:t31i}
\end{table}

Since the charges of the right-handed neutrinos are now bigger than
those in solution (B) used in section~\ref{sec:chiral-t1-3}, the
constrains from $N_{\text{eff}}$ are expected to be stronger. 
In fact, a recent detailed phenomenological analysis of the Dirac
radiative seesaw~\cite{Han:2018zcn} includes the case of a right-handed
neutrino with $\nu=4$ and can be fully applied here.
In particular, the restrictions from $N_{\text{eff}}$ for both scalar ($\sigma^a,\eta^a$)
and Dirac fermion ($\psi$) dark matter cases are more important than the
restriction from $Z_{BL}$ searches at the LHC. We remit the reader
there for further details.

It is clear that after SSB of $\operatorname{U}(1)_{B-L}$ both solutions T1-3-E-I ($\alpha=0$)
in section~\ref{sec:chiral-t1-3}  and T3-1-A-I  ($\alpha=0$) reduce to the Dirac radiative seesaw.
Moreover, we can add extra circulating charges in the loop.
In fact, when all the circulating particles in the loop are color octets~\cite{Reig:2018mdk},
we can  have a bound state dark matter
candidate formed by two Dirac color-octet fermions~\cite{DeLuca:2018mzn}.
With our solutions, the extra $Z_2$ symmetry in~\cite{Reig:2018mdk} is
not longer required.

\section{Conclusions}
\label{sec:con}
We found new and simple models for Dirac neutrino masses within an extension of the Standard Model by an  spontaneously broken $\operatorname{U}(1)_{B-L}$ gauge symmetry. Specifically, we studied the minimal chiral realizations, at one-loop, of the dimension-5 total lepton number conserving operator that gives rise to Dirac neutrino masses without imposing extra symmetries. The minimal models contain three or five chiral fields, two of them with the same charges under $B-L$. In the latter case, their charges can be fixed by the requirement to have a dark matter particle in the spectrum. The full particle content as well as the relevant Lagrangian terms were given for each of these models. We also showed that known solutions with vectorlike fermions can be obtained with just the  single symmetry $\operatorname{U}(1)_{B-L}$. These  new models, therefore, can simultaneously accommodate  one-loop Dirac neutrino masses  and dark matter without invoking any discrete symmetries.

\bigskip

\noindent 
NOTE: During the completion  of this work, a related study~\cite{Bonilla:2018ynb} has
appeared. They also present the solution in section~\ref{sec:vectorlike-solutions} (Table~\ref{tab:t31i}) by fixing $r=1/2$.
 
\section*{Acknowledgments}
Work supported by Sostenibilidad-UdeA, and by COLCIENCIAS through the Grants 111565842691 and 111577657253.  D.R thanks Martin Hirsch for illuminating discussions.
O.Z. acknowledges the ICTP Simons associates program and the kind hospitality of the Abdus Salam ICTP where part of this work was done. 

\appendix

\bibliographystyle{apsrev4-1long}
\bibliography{susy}

\end{document}